# Double Step Annealing for the Recovering of Ion Implantation Defectiveness in 4H-SiC DIMOSFET

Massimo Zimbone[1,a*], Nicolo Piluso[2,b], Grazia Litrico[3,c], Roberta Nipoti[4,d],
Riccardo Reitano[5,e], MariaConcetta Canino[4,f], Maria Ausilia Di Stefano[2,g],
Simona Lorenti[2,h] Francesco La Via[1,i]

[1] IMM-CNR, VIII Strada, 5, 95121 Catania, Italy
[2] STMicroelectronics, Stradale Primosole, 50, 95121 Catania, Italy
[3] Laboratori Nazionali del Sud, via S. Sofia, 62, 95123 Catania, Italy
[4] CNR-IMM, Bologna, via Gobetti, 101, I-40129 Bologna, Italy
[5] Department of Physics and Astronomy, University of Catania, Via S Sofia 64, 95123 Catania, Italy
[a] massimo.zimbone@imm.cnr.it, [b] nicolo.piluso@st.com, [c] grazia.litrico@lns.infn.it,
[d] nipoti@bo.imm.cnr.it, [e] riccardo.reitano@ct.infn.it, [f] mariaconcetta@canino@imm.cnr.it,
[g] mariaausilia.distefano@st.com, [h] simona.loreti@st.com, [i] francesco.lavia@imm.cnr.it

**Abstract.** Thermal annealing plays a crucial role for healing the defectiveness in the ion implanted regions of DIMOSFETs (Double Implanted MOSFETs) devices. In this work, we have studied the effect of a double step annealing on the body (Al implanted) and the source (P implanted) regions of such devices. We found that a high temperature annealing (1750°C, 1h) followed by a lower temperature one (1500°C, 4h) is mandatory to achieve low defects concentration and good crystal quality in both the n- and p- type zones of the device.

### Introduction

The ion implantation process plays a pivotal role for the fabrication of 4H-SiC devices in particular in DIMOSFET (Double Implanted MOSFETs) devices, since it determines the ON resistance and the channel mobility values [1,2]. The common drawback of ion implantation is the generation of detrimental disorder in lattice crystal that needs to be healed to ensure the reliability of the device. The typical process used to decrease the defectiveness of the ion implantation and increase the crystal quality is the annealing at high temperature. Nevertheless, previous studies demonstrated that disorder in the lattice crystal cannot be easily restored by a unique thermal process [3,4]. In particular, some defects can be removed by an appropriate thermal budget, while others can be generated. A fine tuning of thermal processes is indispensable to recover the crystallographic quality of the epilayer and increase the yield and effectiveness of the power device. Hereby, in this work we investigated the defects evolution after ion implantation and double step annealing in 4H-SiC DIMOSFET. The defects evolution, generated by source (P) or body (Al) implantation and subsequent annealing, has been studied in detail by means of micro-photoluminescence (μPL) technique. An annealing at high temperatures (in the range 1650 - 1750°C, 1h) and a subsequent annealing at lower temperature and for longer time (1500°C, 4h and 14h) has been used in order to control both extended and point defects.

### Experimental Setup

The epitaxial layers have been grown in a low-pressure hot-wall chemical vapor deposition reactor by Tokyo Electron Limited (TEL). The epitaxial layers (N doped $10^{16}$ at/cm$^3$) were grown on (001) 4H-SiC n-type (~ $10^{18}$ at/cm$^3$) substrates with 4° off axis. The epilayer thickness was 6 microns. The epitaxies were implanted with typical doses used for DIMOSFET applications: $10^{13}$ - $10^{14}$ cm$^{-2}$ of P+ ions in the source region and $10^{12}$ - $10^{13}$ cm$^{-2}$ of Al+ ions in the body region with an implantation energy of several keV. After the implantation, we performed two thermal annealing procedures.

The first procedure was performed in temperature range of 1650°C to 1750° for 1h while the second was performed at 1500°C for 4 or 14 hours. Samples were characterized by photoluminescence (PL) measurements using a NanoLog Horiba spectro-fluorometer instrument. In order to avoid the contribution from the substrate in the PL, the excitation wavelength was chosen in the deep UV (266 nm), having a penetration length of less than 1 micron in 4H-SiC.

**Results**
**Results and Discussion**
Figures 1(a) and 1(b) show the PL spectra of ion implanted specimens after a first-high temperature annealing. P doped (n-type) samples are shown in figure 1(a) while Al doped (p-type) samples are shown in figure 1(b). Three luminescence bands are apparent: at 388 nm, 485 nm and 762 nm in wavelength. The signal at 388nm is related to band-to-band recombination convoluted to nitrogen related emission. The 485 nm PL band is related to crystallographic damage being present in both implants (Al and P) and being roughly proportional to the implant dose (not shown here). The band at 762 nm, observable in p-type material, is attributed to the implanted species since it appears only in the Al doped samples[3]. Small peak at 532 nm is the second order of the exciting wavelength (266 nm). A decrease of all PL signals is observed with increasing annealing temperature of the first annealing from 1650°C to 1750°C. In particular, the band at 762 nm completely disappears at 1750°C while the bands at 476 nm and 388 nm still linger. In figure 2, the spectra of n-type (figure 2a) and p-type (figure 2b) samples subjected to a double annealing treatment (1650°C 1 h + 1500°C 4h and 14h) are shown. The second (low temperature) annealing induces a decrease of the luminescence of the 476 nm band in both n-type and p-type samples and an increase of PL intensity of the 388 nm (in both n- and p-type) and 762 nm (only in p-type) bands. In figure 3, we report the peak intensity at 388 nm (figure 3a) and 476 nm (figure 3b) of the n-type implanted zone, after and before the low temperature annealing, as a function of the first (high) temperature annealing process. As can be observed from these figures, the low temperature annealing has negligible (within the experimental errors) influence on the 388 nm band whereas a clear decrease of the intensity isnoticeable for 476 nm band. We can infer that the second annealing does not influence the b-b transition related to 388 nm band while reduces the quantity of the defects that produce the luminescence at 476 nm. Moreover it is worth nothing that the intensity of samples annealed for 4h and 14 h has the same value, suggesting that 4 hours are enough to reach defects equilibrium and to complete the kinetics.The behavior of p-type (Al doped) samples is more articulate. The peaksintensity at 388 nm (figure 4a), 476 nm (figure 4b) and 762 nm (figure 4c) as a function of the temperature of the first annealing are shown in figure 4. Second annealing induces an increase of the PL intensity in the 388 and in 762 nm bands and a decrease of the intensity of 476 nm band. It is worth nothing that the first annealing at 1750°C reduces the intensity of 762 nm peak to a negligible value so the increase in intensity (due to the second annealing) is observable only for those samples annealed at temperature lower than 1700°C. As stated before the 388 nm peak is related to b-b transition and its intensity is related to the amount of free carrier concentration. The intensity increase of this peak may be due to an improvement of the crystal quality. This consideration is supported by lifetime measurements (not hown here) which are linearly correlated to the 388 nm PL intensity. The higher carrier concentration (and higher lifetime) induces an increase of the luminescence of the 388 nm and 762 nm peak (if it is present in the sample) as is observed in the figure 4c. Nevertheless, the intensity of 476 nm peak decreases, despite of the behavior of the other two peaks. This behavior clearly indicates that the 476 nm induced defects decrease in concentration. Moreover, as in the n-type material, kinetics is completed in 4 hours.

**Conclusions**
In the present paper, we performed a double step annealing on the body (Al implanted) and source (P implanted) 4H-SiC and observed that the second (low temperature) annealing reduces the defect concentration, improving the crystal quality. In particular, defects related to crystallographic damage (with a PL emission at 472 nm) decrease in concentration whereas the 388 nm band (that is related to the crystal quality) increases in intensity. Aluminum related defects (with a PL emission at 762 nm) can be eliminated. with the first high temperature annealing.

**Acknowledgments**

This work was carried out in the framework of the ECSEL JU project WInSiC4AP (Wide Band Gap Innovative SiC for Advanced Power), grant agreement n. 737483.

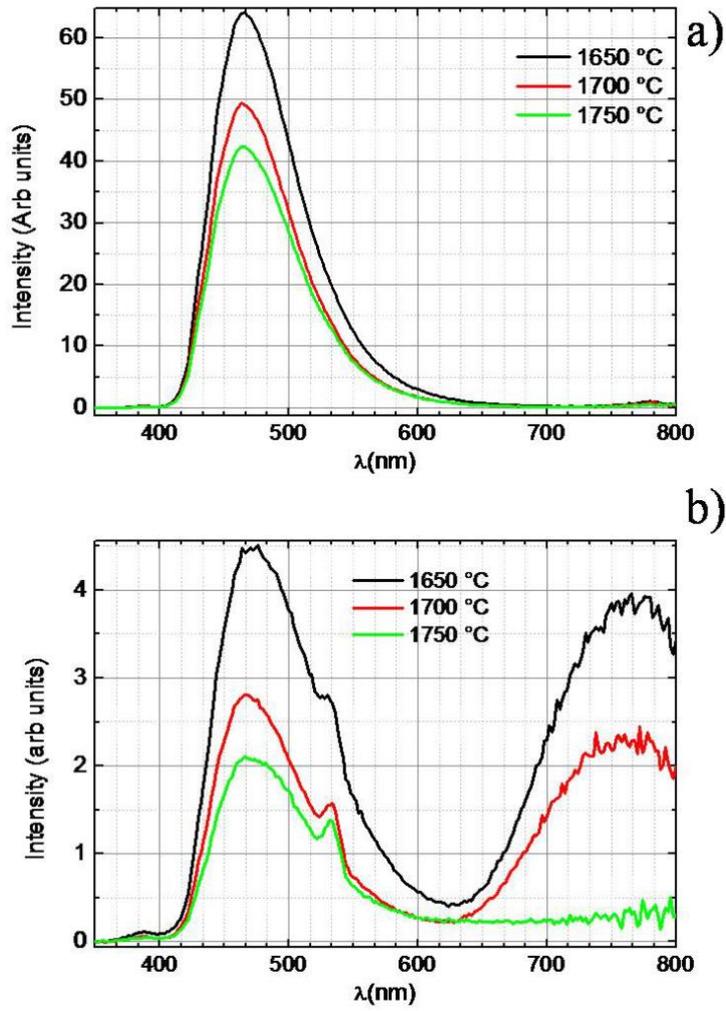

Figure 1: Photoluminescence spectra of P (a) and Al (b) implanted 4H-SiC samples irradiated andd annealed at different annealing temperatures.

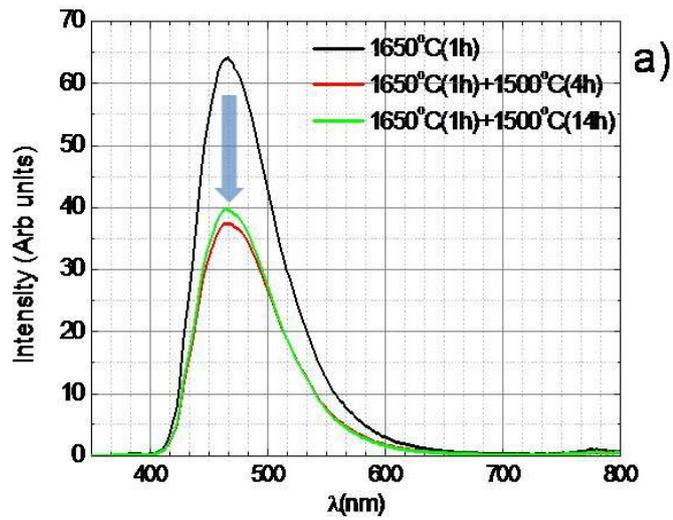

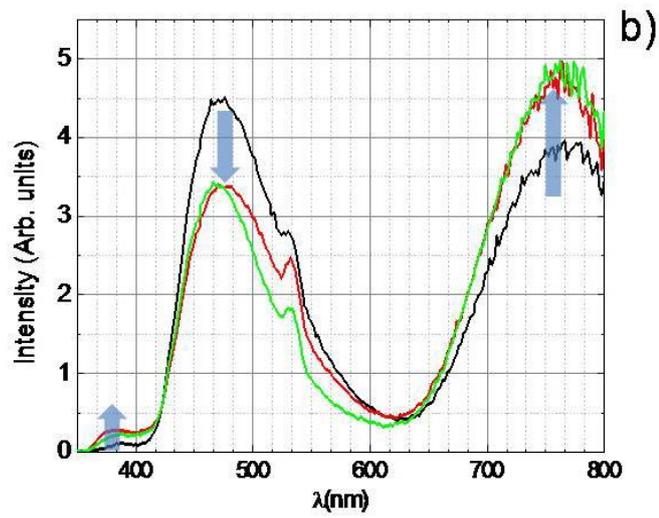

Figure 2: Spectra of n-type P implanted (a) and ptype Al implanted (b) samples annealed at 1650 $_0$C and subsequent annealed at 1500$_o$C for 4h or 14 h. Light blue arrows are drawn to clearly explain the annealing effects

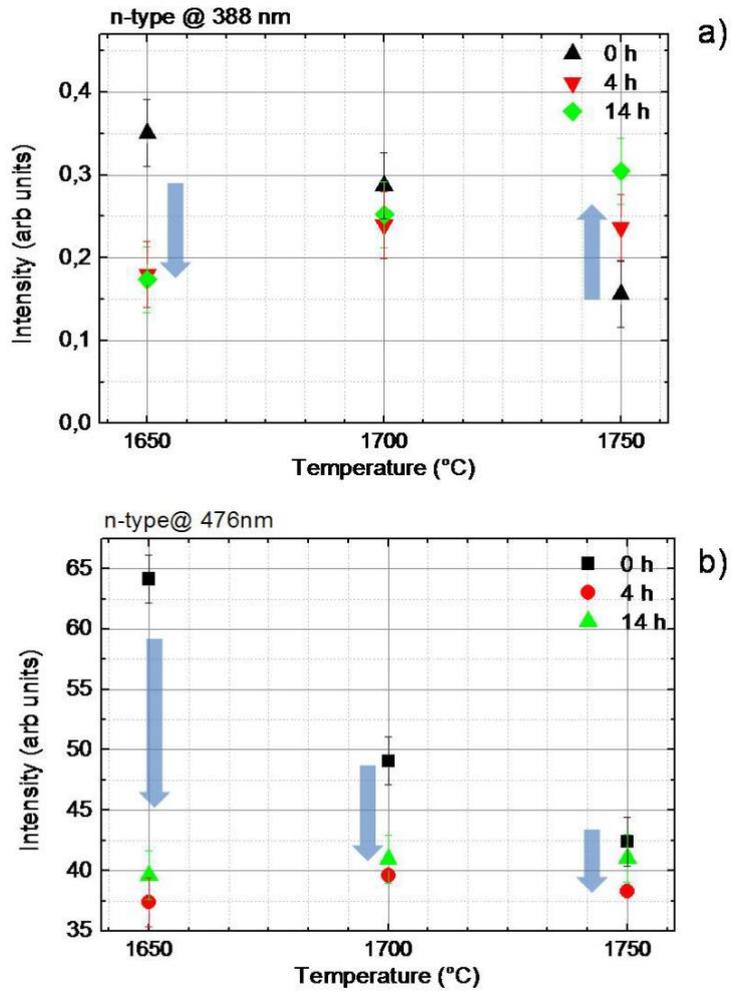

Figure 3: Peak intensity at 388 nm (a) and 476 nm (b) from P implanted samples, after and before the low temperature annealing. Intensities are plotted as a function of the first annealing temperature.

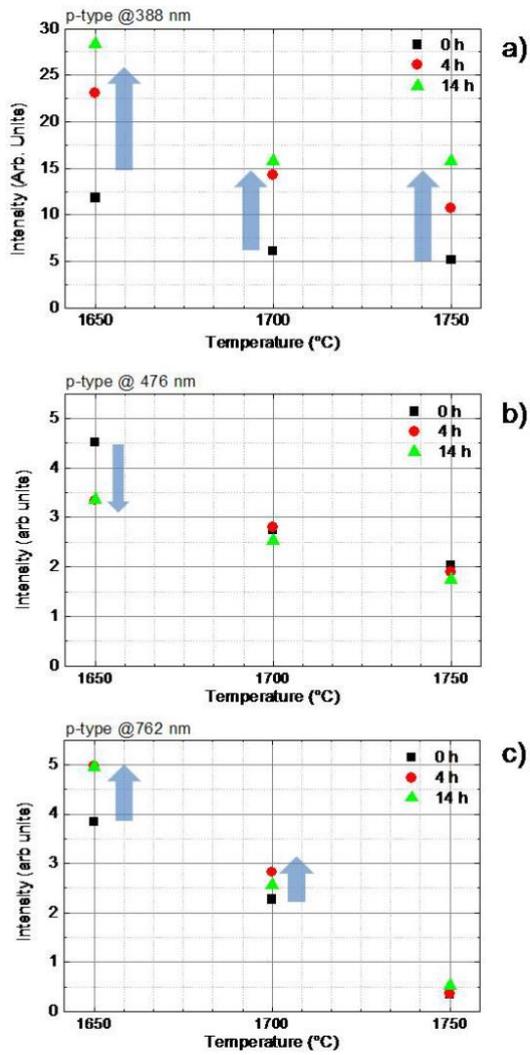

Figure 4: The peak intensity at 388 nm (a), 476 nm (b) and 762 nm (c) from Al implanted zone (ptype 4H- SiC) before and after the low temperature annealing as a function of the first annealing temperature.